# Secret-Key-Aided Scheme for Securing Untrusted DF Relaying Networks

Ahmed El Shafie[†], Ahmed Sultan[‡], Asma Mabrouk[*], Kamel Tourki[⋆], Naofal Al-Dhahir[†]

[†]Electrical Engineering Dept., University of Texas at Dallas, (e-mail: {ahmed.elshafie, aldhahir}@utdallas.edu).
[‡]King Abdullah University of Science and Technology (KAUST), (e-mail: salatino@alumni.stanford.edu).
[*]HANA Research Lab, ENSI, Manouba University, (e-mail: mabroukasma89@gmail.com).
[⋆]Mathematical and Algorithmic Sciences Lab, France Research Center, Huawei Technologies Co. Ltd, (e-mail: kamel.tourki@gmail.com).

*Abstract*—This paper proposes a new scheme to secure the transmissions in an untrusted decode-and-forward (DF) relaying network. A legitimate source node, Alice, sends her data to a legitimate destination node, Bob, with the aid of an untrusted DF relay node, Charlie. To secure the transmissions from Charlie during relaying time slots, each data codeword is secured using a secret-key codeword that has been previously shared between Alice and Bob during the perfectly secured time slots (i.e., when the channel secrecy rate is positive). The secret-key bits exchanged between Alice and Bob are stored in a finite-length buffer and are used to secure data transmission whenever needed. We model the secret-key buffer as a queueing system and analyze its Markov chain. Our numerical results show the gains of our proposed scheme relative to benchmarks. Moreover, the proposed scheme achieves an upper bound on the secure throughput.

## I. INTRODUCTION

Physical layer security (PLS) is a promising layer of defense envisioned to meet the advanced requirements of future communication systems with confidentiality guarantees. The use of secret keys to enhance the security has been adopted in many works [1]–[3]. The authors of [1] proposed the idea of using a key queue in a single-user system where the queue is kept at both the legitimate source, Alice, and the legitimate destination, Bob, and its content is perfectly hidden from the eavesdropping node, Eve. The key idea of this approach is to use a portion of the secrecy rate of the legitimate channel to send randomly-generated key bits. The stored key bits at the secret-key queue can be used later to achieve secure communication between Alice and Bob when the Alice-Bob link is not secured (i.e., when the link's instantaneous secrecy rate is lower than the target secrecy/data rate). Based on this promising idea, the authors of [2] investigated the wiretap channel and showed that a constant long-term secrecy rate is achievable and they further addressed the queueing delay problem without considering random data arrivals at Alice. In [3], under fixed data and key packet rates, the authors proposed two schemes to secure the transmissions of Alice when she is equipped with a data buffer and investigated the data queueing delays.

Unlike all the above-mentioned works, we consider an untrusted DF relaying network. Despite being a legitimate entity in the network and willing to faithfully carry out the designed relaying scheme, the relay node may have a lower security clearance than the legitimate nodes and, hence, is untrusted with the confidential information it is relaying. That is, a cooperative untrusted relay is defined as a relay which is trusted at the service level and untrusted at the data level. Therefore, it should not be able to decode the messages between the source and destination. Hence, it has been stated that it is not possible to ensure secure decode-and-forward (DF) relay-assisted transmissions [4]. Consequently, the amplify-and-forward (AF) is the most commonly used relaying protocol in untrusted cooperative communications. The results in [5] showed that the untrusted AF relay node has better channel conditions to the source than the destination and, hence, can successfully decode the transmitted information.

Cooperative jamming techniques have been widely used in untrusted AF relaying systems [6]–[9]. The secrecy performance of destination-based jamming techniques in two-hop untrusted relay system was analyzed in [6]. In [7], to keep its information secret from the relay, the source allocates a part of its transmit power to send a jamming signal. In [8], joint destination-based jamming and precoding at both the source and the relay was studied for a multi-antenna untrusted AF relay system. Unlike the works where the destination node sends a jamming signal as in, e.g., [8] and the references therein, we assume that the destination can share a secret-key with the source node since this can increase security due to the ability of the keys to achieve the link rate (instead of the secrecy rate). In the presence of the direct source-destination link, the authors in [10] proposed an opportunistic transmission scheme based on the achievable secrecy capacity for AF untrusted relay network. It was pointed out that employing the untrusted relay is useless when the target secrecy rate was larger than a certain threshold value.

Different from the aforementioned works, we propose a secret-key-aided scheme to secure a dual-hop communication network with an untrusted DF relay. The untrusted relay node participates in the transmission only when the Alice-Bob link is not secure. In this case, secrecy can be guaranteed if the legitimate nodes have exchanged enough secret-key bits without being eavesdropped on by the relay node. We consider the general case of full-duplex (FD) Bob and use of limited-size key queues at both Alice and Bob. The stored key bits will be used to secure data transmissions whenever needed. We model the secret-key buffer as a queueing system and analyze its Markov chain (MC). We derive new results for the achievable rates of FD nodes under a block-fading self-interference channel model, where the self-interference channel is parameterized by a block-fading channel model that can be used to study all residual self-interference (RSI) cases including slow-RSI and fast-RSI models. The presented ideas to derive the achievable rate of the Alice-Bob link can be used to derive the rates when the destination is an FD node that aids with cooperative jamming (i.e., sends artificial noise signals) to confuse the passive eavesdropping nodes as in, e.g., [11].

*Notation:* Unless otherwise stated, lower- and upper-case bold letters denote vectors and matrices, respectively. $\mathcal{CN}(x,y)$ denotes a complex circularly-symmetric Gaussian random variable with mean $x$ and variance $y$. $\mathbf{I}_N$ denotes the identity matrix whose size is $N \times N$. $\mathbb{C}^{M \times N}$ denotes the set of all complex matrices of size $M \times N$. $(\cdot)^\top$, $(\cdot)^*$, and $(\cdot)^H$ denote the transpose and Hermitian (i.e., complex-conjugate transpose) operations, respectively. $\lfloor \cdot \rfloor$ denotes the floor of the value in brackets. $|\cdot|$ denotes the absolute of the value in brackets and $\mathbb{E}\{\cdot\}$ denotes statistical expectation. $\mathbf{0}_{M \times N}$ and $\mathbf{1}_{M \times N}$ denote the all-zero and all-ones matrices, respectively, with size $M \times N$. $\mathrm{diag}=\{\cdot\}$ denotes a diagonal matrix with the enclosed elements as its diagonal elements. $[\cdot]^+ = \max\{\cdot, 0\}$ denotes the maximum between the argument and zero.

## II. SYSTEM MODEL AND PROPOSED SCHEME

### A. System Model

We consider a two-hop wireless network with a source node (Alice), an untrusted relay node (Charlie), and an FD legitimate receiver (Bob). Alice tries to send her data to Bob without any information leakage at Charlie. Since Charlie is a part of the network, his CSI is available at both Alice and Bob. We assume a direct link between Alice and Bob. All nodes are assumed to be equipped with single antennas. Both Alice and Bob maintain a secret-key buffer (queue) to store the exchanged keys. Once a key is used, it leaves the buffer. This buffer is assumed to be size limited. Let $L_{\max}$ denote the maximum buffer size, and $Q_A$ denote the number of key bits queued in the buffers of Alice and Bob at the beginning of the transmission slot. We assume that the time is partitioned into slots where each slot has a duration of $T$ and the channel bandwidth is $W$. A data packet is assumed to contain $b$ bits. Hence, the spectral efficiency of a data packet transmission is $\mathcal{R}_d = b/(WT)$. To secure a data packet of size $b$ bits, a key of size $b$ should be used. Hence, the minimum value of $L_{\max}$ is $b$ bits. The transmit power of node $i$ is $P_i$. In the mathematical notation, we use the subscript A, B, and R to denote Alice, Bob, and the relay (Charlie), respectively.

We assume a quasi-static flat-fading channel model where the channel coefficients remain constant during a coherence/slot time duration, but change independently from one coherence/slot time duration to another. The channel coefficient from $i$, $(i \in \{A, B, R\})$, to $j$, $(j \in \{A, B, R\})$, is denoted by $h_{ij}$ and distributed as $\mathcal{CN}(0, \sigma_{ij}^2)$. The channel gain, squared-value of the channel coefficient, of the $i-j$ link is $g_{ij}$. We assume channel reciprocity so that $h_{ij} = h_{ji}$. The thermal noise at node $k$ is modeled as additive white Gaussian noise (AWGN) with zero mean and power $\kappa_k$.

Even when the wireless channels are not experiencing fading, the RSI channel is time-varying [12], [13]. The instantaneous variations of the RSI channel are due to the cumulative effects of various distortions originating from noise, carrier frequency offset, oscillator phase noise, analog-to-digital/digital-to-analog (AD/DA) conversion imperfections, in-phase/quadrature (I/Q) imbalance, imperfect channel estimation, etc. These impairments are inevitable and cannot be completely eliminated. In addition, they have a notable effect on the RSI channel due to the very close distance between the transmitter-end and the receiver-end of the self-interference (SI) channel. Furthermore, since the variations are random, they cannot be accurately estimated at the FD terminal [12]. Unlike most of the literature which assumes slow-RSI model[1] that captures only the long-term, i.e., codeword-by-codeword, statistical properties of the RSI channel, we assume the general case of block-fading RSI channel. As explained in Appendix A, this model is parameterized with a block-fading parameter $M$ with $M=1$ for the fast-RSI channel and $M \to \infty$ for the slow-RSI channel.

## III. PROPOSED SCHEME AND RATE EXPRESSIONS

In our proposed scheme, there are two possible transmission modes: (1) the direct transmission (DT) mode, where Alice sends the data to Bob over the entire time slot duration without the need for Charlie, or (2) the relaying transmission (RT) mode, where the time slot is divided into two non-overlapping phases and Charlie decodes and forwards Alice's transmission. When the Alice-Bob link is not secured without using a secret key packet, the RT mode is favorable over the DT mode since, in addition to data, it allows for extra key bits sharing between Alice and Bob by virtue of Bob's FD capability as explained below. Hence, the RT mode is used whenever the achievable end-to-end (e2e) rate of the Alice-Bob link is higher than $\mathcal{R}_d$. In both transmission modes, Alice uses a secret key with rate $\mathcal{R}_d$ since Alice transmits one data packet with that size. During the second phase of the RT mode, Alice is silent, Charlie transmits, and Bob receives. Since Bob is an FD terminal, he can receive data from Charlie and simultaneously send a new key to Alice, who operates in a receive mode. Since Charlie is busy with information forwarding, he will not intercept the key. We note that, as will be shown in Appendix A, in the case of a slow-RSI channel, the self-interference can be completely eliminated at Bob due to the fact that Bob knows what he sends (i.e., knows the transmitted key's codeword prior to transmission). Hence, the legitimate system achieves interference-free rates for all links with no loss. If the data sent from Charlie is in connection outage under the FD transmission mode, the half-duplex (HD) mode is used and only the data will be sent. Note that Charlie causes interference at Alice. However, since Alice knows the signal sent by Charlie (because Alice sent this data signal to Charlie in the first phase), she will be able to remove it prior to data decoding. Accordingly, the received key codeword from Bob is interference free. In the following, we provide some useful definitions that will be used subsequently.

### A. Rate Expressions

In this section, we state the various rates of the wireless links when utilizing our proposed scheme in different communication phases. The rate of the Alice-Bob and Alice-Charlie links in bits/channel use are given, respectively, by

$$\mathcal{R}_{\mathrm{AB}}^{\mathrm{DT}} = \log_2\left(1 + \gamma_{\mathrm{AB}} g_{\mathrm{AB}}\right), \quad \mathcal{R}_{\mathrm{AR}} = \log_2\left(1 + \gamma_{\mathrm{AR}} g_{\mathrm{AR}}\right) \quad (1)$$

---
[1]Typically, slow-RSI model is assumed in the literature (e.g., [11], [14], [15]), which represents an optimistic assumption.

where $\gamma_{ij} = \frac{P_i}{\kappa_j}$. If the DT mode is utilized, the secrecy rate using the entire time slot length is

$$\mathcal{R}_{\text{AB}}^{\text{sec}} = [\mathcal{R}_{\text{AB}}^{\text{DT}} - \mathcal{R}_{\text{AR}}]^+ \quad (2)$$

Similarly, when Bob sends a key to Alice, the secret-key length, which is equal to the secrecy rate of the Bob-Alice link, is given by

$$\mathcal{R}_{\text{BA}}^{\text{sec}} = [\mathcal{R}_{\text{BA}} - \mathcal{R}_{\text{BR}}]^+ \quad (3)$$

We emphasize here that the secrecy rate expression of the Alice-Bob link is *not equal* to the secrecy rate expression of the Bob-Alice link because the rate of the Bob-Charlie (eavesdropping's) link is different from the rate of the Alice-Charlie link since the two links experience different channel realizations.

If Charlie is asked to help Alice, the e2e data rate of the Alice-Bob link when Bob does not send a key (i.e., operates in an HD receiving mode) is given by

$$\mathcal{R}_{\text{AB}}^{\text{RT,HD}} = 0.5 \min\left\{\mathcal{R}_{\text{AR}}, \log_2\left(1 + g_{\text{RB}}\gamma_{\text{RB}} + g_{\text{AB}}\gamma_{\text{AB}}\right)\right\} \quad (4)$$

If Bob decides to send a key while receiving Charlie's transmission, the achievable e2e data rate of the Alice-Bob link, $\mathcal{R}_{\text{AB}}^{\text{RT}}$, when he combines the two received signals from Alice (during the first phase) and from Charlie (during the second phase) is given by (14) in Appendix A. We assume that Bob adjusts his power level to make the e2e rate of the Alice-Bob link greater than the target data rate, $\mathcal{R}_d$. That is, Bob's transmit power is set to equal to its maximum value that makes $\mathcal{R}_{\text{AB}}^{\text{RT}} \geq \mathcal{R}_d$. By doing this, the HD mode is guaranteed to be used when there is no positive power level that can be used by Bob to achieve the target rate $\mathcal{R}_d$. Moreover, selecting the maximum value will allow Bob to send more secret-key bits to Alice.

### B. Proposed Scheme

Our proposed scheme is summarized as follows in order:
- If the instantaneous secrecy rate of the DT mode supports the secure transmission rate, Alice sends a data packet to Bob with rate $\mathcal{R}_d$. If $\mathcal{R}_{\text{AB}}^{\text{sec}} > \mathcal{R}_d$, key bits with rate $\mathcal{R}_{\text{AB}}^{\text{sec}} - \mathcal{R}_d$ are also transmitted.
- If $\mathcal{R}_{\text{AB}}^{\text{sec}} < \mathcal{R}_d$, depending on the secret-key queue state, we distinguish here among three cases.
  – If the secret key queue has no sufficient bits to secure a data packet (i.e., its state is less than $\mathcal{R}_d$ bits/sec/Hz), either Alice sends a key to Bob or Bob sends a key to Alice based on the instantaneous secrecy rates of the Alice-Bob link and the Bob-Alice link. That is, a secret key with rate $\max\{\mathcal{R}_{\text{AB}}^{\text{sec}}, \mathcal{R}_{\text{BA}}^{\text{sec}}\}$ should be shared between Alice and Bob.
  – If the secret-key queue is not full but has sufficient bits to secure a data packet (i.e., its state is at least $b$ bits), Alice checks whether the rate of the Alice-Bob or Alice-Charlie-Bob links (not secrecy rate) is not in connection outage (i.e., the rate is higher than $\mathcal{R}_d$). If any of the two links is not in connection outage, Alice sends a data packet to Bob either directly or through Charlie after XORing it with key bits (which is known as the one-time pad technique). Otherwise, either Alice sends a key to Bob or Bob sends a key to Alice based on the instantaneous secrecy rates of the Alice-Bob link and the Bob-Alice link. That is, a secret key with rate $\max\{\mathcal{R}_{\text{AB}}^{\text{sec}}, \mathcal{R}_{\text{BA}}^{\text{sec}}\}$ should be shared between Alice and Bob.
  – If the secret-key queue is full of key bits, Alice and Bob cannot share/exchange any new key bits. Alice checks whether the rate of the Alice-Bob or Alice-Charlie-Bob links is not in connection outage. If both links are in connection outage, all nodes remain silent. Otherwise, Alice sends a data packet to Bob either directly or through Charlie after XORing it with key bits.

It should be emphasized here that if the number of transmitted key bits exceeds the buffer size, the extra bits will be discarded at both Alice and Bob buffers.

### C. Secret-Key MC

We model the secret-key buffers as an MC where each state represents the number of bits in the queue. Let $\mathcal{T}^m$ and $\mathcal{T}^{-m}$ denote the transition probabilities that the queue size increases and decreases by $m$ bits, respectively. Let $\zeta_b$ denotes an indicator which is equal to one if at least $b$ bits are stored in the queue and zero otherwise. In particular, if the queue is full, $\zeta_b = 1$. To simplify the notation, we define a function $\tilde{R} = \lfloor WTR \rfloor$ to represent the number of bits in a transmission over bandwidth $W$ and time duration $T$ with rate $R$ bits/sec/Hz. Based on the proposed description, during any time slot, each buffer status can be modified as follows

- *The non-full queue remains unchanged:* The queue size remains unchanged in three cases: 1) if the DT mode is secure and can support only $b$ bits transmissions, 2) if the FD mode is used and $b$ key bits are exchanged between Alice and Bob; or 3) in case of secrecy outage (i.e., all the nodes remain silent). In this case, the transition probability is given by

$$\begin{aligned}\mathcal{T}^0 &= \Pr\{\tilde{R}_{\text{AB}}^{\text{sec}} \geq b, \tilde{R}_{\text{AB}}^{\text{sec}} < b+1\} \\ &+ \zeta_b \Pr\{\tilde{R}_{\text{AB}}^{\text{sec}} < b, \tilde{R}_{\text{AB}}^{\text{RT}} \geq b, b \leq 0.5\tilde{R}_{\text{BA}}^{\text{sec}} < b+1\} \\ &+ \Big[(1-\zeta_b)\Pr\{\tilde{R}_{\text{AB}}^{\text{sec}} = 0, \tilde{R}_{\text{BA}}^{\text{sec}} = 0\} \\ &+ \zeta_b \Pr\{\max\left(\tilde{R}_{\text{AB}}^{\text{DT}}, \tilde{R}_{\text{AB}}^{\text{RT}}\right) < b, \tilde{R}_{\text{AB}}^{\text{sec}} = 0, \tilde{R}_{\text{BA}}^{\text{sec}} = 0\}\Big] \end{aligned} \quad (5)$$

- *The queue remains full:* With full queue, Alice and Bob cannot store more key bits. Moreover, if both DT and RT transmission modes do not support the transmission rate $b$, they remain silent.

$$\begin{aligned}\mathcal{T}^0 &= \Pr\{\tilde{R}_{\text{AB}}^{\text{sec}} < b, \tilde{R}_{\text{AB}}^{\text{RT}} \geq b, b \leq 0.5\tilde{R}_{\text{BA}}^{\text{sec}} < b+1\} \\ &+ \Pr\{\tilde{R}_{\text{AB}}^{\text{sec}} < b, \max\left(\tilde{R}_{\text{AB}}^{\text{DT}}, \tilde{R}_{\text{AB}}^{\text{RT}}\right) < b\}\end{aligned} \quad (6)$$

- *The non-full queue size increases:* As long as the queue is not full, some key bits could be shared between Alice and Bob. The probability that the queue size increases with $m$ bits is given by (7) at the top of the next page.

$$\mathcal{T}^m = \Pr\{\tilde{R}_{\text{AB}}^{\text{sec}} - b \geq m, \tilde{R}_{\text{AB}}^{\text{sec}} - b < m+1\} + \zeta_b \Pr\{\tilde{R}_{\text{AB}}^{\text{sec}} < b, \tilde{R}_{\text{AB}}^{\text{RT}} \geq b, 0.5\tilde{R}_{\text{BA}}^{\text{sec}} \geq b+m, 0.5\tilde{R}_{\text{BA}}^{\text{sec}} < b+m+1\}$$
$$+ \left[(1-\zeta_b)\Pr\{\tilde{R}_{\text{AB}}^{\text{sec}} < b, m \leq \max\left(\tilde{R}_{\text{AB}}^{\text{sec}}, \tilde{R}_{\text{BA}}^{\text{sec}}\right) < m+1\} + \zeta_b \Pr\left\{\max\left(\tilde{R}_{\text{AB}}^{\text{DT}}, \tilde{R}_{\text{AB}}^{\text{RT}}\right) < b, \tilde{R}_{\text{AB}}^{\text{sec}} < b, m \leq \max\left(\tilde{R}_{\text{AB}}^{\text{sec}}, \tilde{R}_{\text{BA}}^{\text{sec}}\right) < m+1\right\}\right] \quad (7)$$

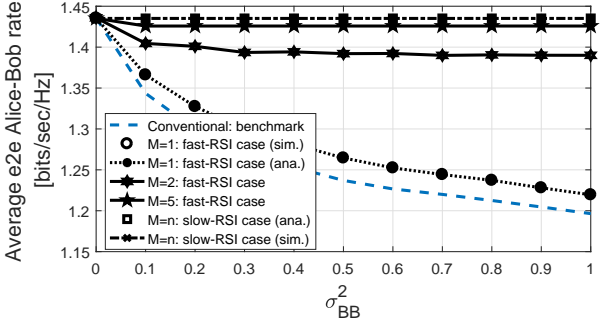

Fig. 1. Average e2e rate of the Alice-Bob link.

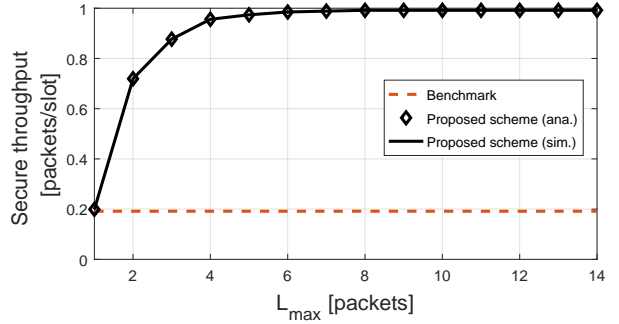

Fig. 2. Throughput in packets/slot versus maximum secret-key buffer size.

- *The non-empty queue size decreases:* The queue size is decreased only when the secret-key queue contains sufficient bits to secure a data packet and either the DT or RT mode is not in connection outage. If either the DT or HD-RT mode is used, the queue size decreases by $b$ bits. In this case, the transition probability can be evaluated as

$$\mathcal{T}^{-b} = \zeta_b \Pr\{\tilde{R}_{\text{AB}}^{\text{sec}} < b, \tilde{R}_{\text{AB}}^{\text{RT}} \geq b, \tilde{R}_{\text{BA}}^{\text{sec}} = 0\} \\ + \zeta_b \Pr\{\tilde{R}_{\text{AB}}^{\text{sec}} < b, \tilde{R}_{\text{AB}}^{\text{RT}} < b, \tilde{R}_{\text{AB}}^{\text{DT}} \geq b\} \quad (8)$$

On the other hand, if FD-RT mode is used, $b$ bits leaves the queue to secure the data transmission while Bob shares some key bits with Alice in the second phase. In this case, if Bob sends $m$ key bits ($m < b$), the transition probability is given by

$$\mathcal{T}^{-(b-m)} = \zeta_b \Pr\{\tilde{R}_{\text{AB}}^{\text{sec}} < b, \tilde{R}_{\text{AB}}^{\text{RT}} \geq b, m \leq 0.5\tilde{R}_{\text{BA}}^{\text{sec}} < m+1\} \quad (9)$$

After constructing the MC transition probabilities matrix, denoted by $\mathbf{T}_A$, the steady-state distribution vector for the given transitions is computed by calculating higher powers of the matrix [16]

$$\boldsymbol{\pi} = \boldsymbol{\pi}^{(0)} \times (\mathbf{T}_A)^\infty \quad (10)$$

since the MC is irreducible with ergodic (i.e., aperiodic positive-recurrent) states. Note that all initial distributions, $\boldsymbol{\pi}^{(0)}$, will eventually lead to the same steady-state [16].

The secure throughput in packets/slot is given by

$$\mu_{\text{sec}} = \Pr\{\mathcal{R}_{\text{AB}}^{\text{sec}} \geq \mathcal{R}_d\} \\ + \Pr\{Q_A \geq b\} \bigg( \Pr\{\mathcal{R}_{\text{AB}}^{\text{sec}} < \mathcal{R}_d, \mathcal{R}_{\text{AB}}^{\text{RT}} \geq \mathcal{R}_d\} \\ + \Pr\{\mathcal{R}_{\text{AB}}^{\text{sec}} < \mathcal{R}_d, \mathcal{R}_{\text{AB}}^{\text{RT}} < \mathcal{R}_d, \mathcal{R}_{\text{AB}}^{\text{DT}} \geq \mathcal{R}_d\} \bigg) \quad (11)$$

## IV. SIMULATION RESULTS

We assume that each channel coefficient is a complex Gaussian random variable with zero mean and unit variance. The parameters used to generate the figures are $P_A = 10$ dBm, $WT = 1000$, $b = 2000$ bits, $P_B = P_R = 20$ dBm, $\kappa_k = \kappa = 0$ dBm (for all $k$), and $\sigma_{\text{BB}}^2 = 0.2$. In Fig. 1, we evaluate the average e2e achievable rate of the Alice-Bob link when the relay is utilized. The figure shows that the new proposed achievable rates are higher than those proposed in the literature for FD communications which ware evaluated under the impractical assumption of known RSI channel coefficient and unknown data. In our approach, we exploit the fact that Bob knows the transmitted key (since he randomly generates it) and hence can do better in decoding the received signal from the relay node. The figure also verifies the closed-form expressions for the achievable rates in the fast-RSI case given in Eqn. (19) and the slow-RSI case given in Eqn. (21).

In Fig. 2, we plot the secure throughput versus the maximum buffer size $L_{\max}$ in packets, where a packet is composed of $b = 2000$ bits and $P_A = 20$ dBm. The secure throughput is a non-decreasing function of $L_{\max}$. For the given system's parameters, $L_{\max} \geq L_o = 7$ packets achieves the same secure throughput when $\mathcal{R}_d = 2$ bits/sec/Hz. This means that Alice and Bob can make the buffer maximum size equal to 7 without loss of optimality. As discussed in the system model, $L_{\max} \geq b$ bits (2 packets) is necessary for the proposed scheme to be effective. The minimum Fig. 2 also shows the gain of relaying using an untrusted relay relative to the benchmark that the relay is not utilized. It can be shown that the upper bound that 1 data packet per time slot is securely received at Bob is attained under our proposed scheme. We emphasize that this upper bound is an upper bound on both throughput without eavesdropping and secure throughput.

## V. CONCLUSIONS

We showed that DF relaying can be used for untrusted relaying networks. We derived closed-form expressions for the achievable rates. We showed that increasing the buffer size of the secret-key queue can increase the achievable secure throughput. We also showed that, for a given target secrecy rate, there is a limited buffer size $L_o$ such that the secure throughput will become saturated (constant). All our analytical findings have been verified numerically.

## APPENDIX A
### THE E2E ACHIEVABLE RATE OF THE ALICE-BOB LINK

We derive the e2e achievable rate of the Alice-Bob link using the same approach in [13] since Bob knows his transmitted signal/codeword. The received signal at Bob during the second phase is

$$\mathbf{y}_B = [y_B(1),\ldots,y_B(n)]^\top = h_{RB}\mathbf{x}_R + \text{diag}\{\mathbf{x}_B\}\mathbf{h}_{BB} + \mathbf{z}_B \quad (12)$$

where $n$ is number of symbols in a codeword, $\mathbf{x}_R = [x_R(1),\ldots,x_R(n)]^\top$, $x_R(t)$ is the relay's transmit signal at symbol duration $t$, $h_{BB}(t)$ is the RSI channel coefficient at symbol duration $t$, $\mathbf{h}_{BB} = [h_{BB}(1),\ldots,h_{BB}(n)]^\top$, $\mathbf{x}_B = [x_B(1),\ldots,x_B(n)]^\top$, $x_B(t)$ is Bob's transmit signal at symbol duration $t$, $\mathbf{z}_B = [z_B(1),\ldots,z_B(n)]^\top$, and $z_B(t)$ is the AWGN signal at symbol duration $t$. Since the vector $\mathbf{x}_B$ is known at Bob, the mutual information between $\mathbf{y}_B$ and $\mathbf{x}_R$ should be computed given $\mathbf{x}_B$. Without using this information, if Bob decides to send a key while receiving Charlie's transmission, the achievable e2e data rate of the Alice-Bob link is **conventionally** given by

$$\mathcal{R}_{AB}^{RT} = 0.5 \min\left\{\mathcal{R}_{AR}, \log_2\left(1 + \frac{g_{RB}\gamma_{RB}}{1+g_{BB}\gamma_{BB}} + g_{AB}\gamma_{AB}\right)\right\} \quad (13)$$

which is the widely-used rate expression in the FD communications literature (see, e.g., [11], [15]).

Based on the equation model in (12), and given that Bob knows his transmitted packet $\mathbf{x}_B$ and that he does employ a maximum ratio combiner to decode both received data signals from Alice and Charlie, the e2e achievable rate of the Alice-Bob link, denoted by $\mathcal{R}_{AB}^{RT}$, is given by

$$\begin{aligned}\mathcal{R}_{AB}^{RT} =& \frac{1}{2}\log_2\left(1 + |h_{AB}|^2\frac{P_A}{\kappa_B} + |h_{RB}|^2\frac{P_R}{\kappa_B}\right) \\ &+ \frac{1}{2n}\log_2\det\left(\gamma_1 \text{diag}\{\mathbf{x}_B\}\mathbf{C}_{\mathbf{h}_{BB}}\text{diag}\{\mathbf{x}_B^*\} + \mathbf{I}_n\right) \\ &- \frac{1}{2n}\log_2\det\left(\gamma_2 \text{diag}\{\mathbf{x}_B\}\mathbf{C}_{\mathbf{h}_{BB}}\text{diag}\{\mathbf{x}_B^*\} + \mathbf{I}_n\right)\end{aligned} \quad (14)$$

where $\mathbf{C}_{\mathbf{h}_{BB}} = \mathbb{E}\{\mathbf{h}_{BB}\mathbf{h}_{BB}^*\}$, $\gamma_1 = \frac{1+|h_{AB}|^2 P_A}{(\kappa_B + |h_{AB}|^2 P_A + |h_{RB}|^2 P_R)}$ and $\gamma_2 = \frac{1}{\kappa_B}$. Next, we evaluate the term $\log_2\det\left(\gamma_q\text{diag}\{\mathbf{x}_B\}\mathbf{C}_{\mathbf{h}_{BB}}\text{diag}\{\mathbf{x}_B^*\} + \mathbf{I}_n\right)$ with $q\in\{1,2\}$.

Assume that the RSI channel is block-faded with block size $M \leq n$. Since the channel is fixed for $M$ consecutive symbol durations (i.e., $\mathbf{h}_{BB}(t+1) = \cdots = \mathbf{h}_{BB}(t+M)$ for $t \in \{0, M, 2M, \ldots\}$), the covariance matrix $\mathbf{C}_{\mathbf{h}_{BB}}$, assuming block-fading, is given by $\mathbf{C}_{\mathbf{h}_{BB}} = \{h_{BB}(1)\mathbf{1}_{M\times M}, h_{BB}(M+1)\mathbf{1}_{M\times M}, \ldots, h_{BB}(\lfloor\frac{n}{M}\rfloor)\mathbf{1}_{M\times M}, h_{BB}(\lfloor\frac{n}{M}\rfloor + 1)\mathbf{1}_{\hat{m}\times\hat{m}}\}$ where $\lfloor n/M\rfloor + 1$ is the number of block per data codeword of length $n$ and $\hat{m} = n - M\lfloor\frac{n}{M}\rfloor$. The second and third terms in the rate expression in (14) can be rewritten as

$$\begin{aligned}\mathcal{X}_q &= \log_2\det\left(\gamma_q\text{diag}\{\mathbf{x}_B\}\mathbf{C}_{\mathbf{h}_{BB}}\text{diag}\{\mathbf{x}_B^*\} + \mathbf{I}_n\right) \\ &= \log_2\det\left(\gamma_q\text{diag}\{\mathbf{x}_B^*\}\text{diag}\{\mathbf{x}_B\}\mathbf{C}_{\mathbf{h}_{BB}} + \mathbf{I}_n\right)\end{aligned} \quad (15)$$

where $q \in \{1,2\}$. The last equality in (15) holds from Sylvester's determinant identity.

The matrix $\mathbf{\Lambda}_{\mathbf{x}_B} = \text{diag}\{\mathbf{x}_B^*\}\text{diag}\{\mathbf{x}_B\} = \text{diag}\{|x_B(1)|^2, |x_B(2)|^2, \ldots, |x_B(n)|^2\}$ is diagonal with $|x_B(t)|^2$ as its $t$-th diagonal entry, and $\mathbf{C}_{\mathbf{h}_{BB}}$ is a block-diagonal matrix. The product of the two matrices results in a block-diagonal matrix with block size $M \times M$. The determinant of a block-diagonal matrix is the product of the determinant of each block. Hence,

$$\begin{aligned}\mathcal{X}_q =& \left(\log_2\left(1 + \gamma_q\sigma_{BB}^2\sum_{t=1}^M |x_B(t)|^2\right)\right. \\ &+ \log_2\left(1 + \gamma_q\sigma_{BB}^2\sum_{t=M+1}^{2M} |x_B(t)|^2\right) + \ldots \\ &\left.+ \log_2\left(1 + \gamma_q\sigma_{BB}^2\sum_{t=n-M\lfloor n/M\rfloor}^n |x_B(t)|^2\right)\right)\end{aligned} \quad (16)$$

### A. Fast RSI

In the case of fast fading where $M = 1$ and the RSI changes each symbol duration, we have the following

$$\mathcal{X}_q = \sum_{t=1}^n \log_2\left(1 + \gamma_q|x_B(t)|^2\right) \quad (17)$$

Since $n$ is very large, from the strong law of large numbers, $\frac{1}{2n}\sum_{t=1}^n \log_2\left(1 + \gamma_q|x_B(t)|^2\right)$ will almost surely converge to $\mathbb{E}\left\{\log_2\left(1 + \gamma_q|x_B(t)|^2\right)\right\}$. Since $|x_B(t)|^2$ is exponentially-distributed random variable, the average of $\log_2\left(1 + \gamma_q|x_B(t)|^2\right)$ is given by

$$\begin{aligned}\mathbb{E}\left\{\log_2\left(1 + \gamma_q|x_B(t)|^2\right)\right\} &= \int_0^\infty \log_2\left(1 + \gamma_q|x_B(t)|^2\right) d|x_B(t)|^2 \\ &= \frac{1}{\ln(2)}\exp\left(\frac{1}{\gamma_q\sigma_{BB}^2}\right)\text{Ei}\left(\frac{1}{\gamma_q\sigma_{BB}^2 P_B}\right)\end{aligned} \quad (18)$$

where $\text{Ei}(x) = \int_x^\infty \frac{\exp(-u)}{u} du$ is the exponential integral. Substituting in the rate expression in (14), we get

$$\begin{aligned}\mathcal{R}_{AB}^{RT} =& \frac{1}{2}\log_2\left(1 + |h_{AB}|^2\frac{P_A}{\kappa_B} + |h_{RB}|^2\frac{P_R}{\kappa_B}\right) \\ &+ \frac{1}{2\ln(2)}\exp\left(\frac{1}{\gamma_1\sigma_{BB}^2}\right)\text{Ei}\left(\frac{1}{\gamma_1\sigma_{BB}^2 P_B}\right) \\ &- \frac{1}{2\ln(2)}\exp\left(\frac{1}{\gamma_2\sigma_{BB}^2}\right)\text{Ei}\left(\frac{1}{\gamma_2\sigma_{BB}^2 P_B}\right)\end{aligned} \quad (19)$$

### B. Slow RSI

In the case of slow fading where $M = n$ and the RSI changes once per codeword, we have the following

$$\log_2\det\left(\gamma_q\mathbf{\Lambda}_{\mathbf{x}_B}\mathbf{C}_{\mathbf{h}_{BB}} + \mathbf{I}_n\right) = \log_2\left(1 + \gamma_q\sigma_{BB}^2\sum_{t=1}^n |x_B(t)|^2\right) \quad (20)$$

where $\frac{1}{2n}\log_2\left(1 + \gamma_q\sigma_{BB}^2\sum_{t=1}^n |x_B(t)|^2\right) = 0$ almost surely when $n \to \infty$. From the strong law of large numbers, $\sum_{t=1}^n |x_B(t)|^2 = n\mathbb{E}\{|x_B(t)|^2\} = nP_B$ when $n$ is large. Taking the limit of (14) when $n$ goes to infinity, $\lim_{n\to\infty}\frac{1}{2n}\log_2\left(\gamma_q\sigma_{BB}^2 nP_B + 1\right) = 0$. Hence, the data rate becomes

$$\mathcal{R}_{AB}^{RT} = 0.5\log_2\left(1 + |h_{AB}|^2\frac{P_A}{\kappa_B} + |h_{RB}|^2\frac{P_R}{\kappa_B}\right) \quad (21)$$

which is the rate of the e2e Alice-Bob channel with no interference. This implies that the RSI can be completely mitigated in the case of slow RSI.